# Very high frequency probes for atomic force microscopy with silicon optomechanics

L. Schwab[1], P. E. Allain[2], N. Mauran[1], X. Dollat[1], L. Mazenq[1], D. Lagrange[1], M. Gély[3], S. Hentz[3], G. Jourdan[3], I. Favero[2] and B. Legrand[1,a]

[1]*Laboratoire d'Analyse et d'Architecture des Systèmes, Université de Toulouse, CNRS UPR 8001, Toulouse, 31031, France*

[2]*Laboratoire Matériaux et Phénomènes Quantiques, Université de Paris, CNRS UMR 7162, Paris, 75013, France*

[3]*Université Grenoble Alpes, CEA, LETI, Minatec Campus, Grenoble, 38000, France*

---

[a] Author to whom correspondence should be addressed. Electronic mail: bernard.legrand@laas.fr
T.: +33 5 6133 6811, F.: +33 5 6133 6300

# Abstract

Atomic force microscopy (AFM) has been constantly supporting nanosciences and nanotechnologies for over 30 years, being present in many fields from condensed matter physics to biology. It enables measuring very weak forces at the nanoscale, thus elucidating interactions at play in fundamental processes. Here we leverage the combined benefits of micro/nanoelectromechanical systems and cavity optomechanics to fabricate a sensor for dynamic mode AFM at a frequency above 100 MHz. This is two decades above the fastest commercial AFM probes, suggesting opportunity for measuring forces at timescales unexplored so far. The fabrication is achieved using very-large scale integration technologies inherited from photonic silicon circuits. The probe's ring optomechanical cavity is coupled to a 1.55 μm laser light and features a 130 MHz mechanical resonance mode with a quality factor of 900 in air. A limit of detection in displacement of $3.10^{-16}$ m/√Hz is obtained, enabling the detection of the Brownian motion of the probe and paving the way for force sensing experiments in the dynamic mode with a working vibration amplitude in the picometer range. Inserted in a custom AFM instrument embodiment, this optomechanical sensor demonstrates the capacity to perform force-distance measurements and to maintain a constant interaction strength between tip and sample, an essential requirement for AFM applications. Experiments show indeed a stable closed-loop operation with a setpoint of 4 nN/nm for an unprecedented sub-picometer vibration amplitude, where the tip-sample interaction is mediated by a stretched water meniscus.

# Introduction

Atomic Force Microscopy (AFM) has many applications in nanosciences, from micro and nanotechnologies to nanobiology. AFM has rapidly become a standard technique for surface observation and force spectroscopy at the nanoscale, the instruments having constantly gained in performance[1,2,3,4,5]. This is particularly true for the dynamic mode, where the probe tip is driven in oscillation close to a mechanical resonance of the probe[6,7]. This mode was progressively preferred to the contact mode as it increases the measurement sensitivity while reducing the degradation of fragile and soft samples like those found in biological experiments[8]. In this oscillating mode, the parameters of the probe, namely mechanical resonance frequency, quality factor and stiffness, have a direct impact on measurement bandwidth, vibration amplitude and force sensitivity. They bound in turn the achievable imaging rate and time and spatial resolution in force tracking. Over the last 15 years, impressive results have been obtained on the way to improve the temporal and spatial resolution in AFM, with great impact in basic science and technology. On one hand, up to tens of frames per second imaging is now achieved, allowing the direct observation of biological processes under native-like conditions[4,5,9]. On the other hand, operation at sub-nanometer amplitude, typically 100 pm, gives access to atomic resolution in vacuum, in air and in liquids[10,11,12]. Molecular bonds imaging has even been achieved in this regime of low vibration[13]. However, such a speed and such extreme resolution cannot be obtained simultaneously. Here, we propose an AFM probe technology based on silicon optomechanics able to overcome the limitation.

The AFM probe is basically a force sensor built from a mechanical resonator. The existing AFM probe technologies are presented in Figure 1. Depending on the vibration mode, dimension and mechanical properties of the resonator, different ranges of frequency and vibration amplitude may be reached.

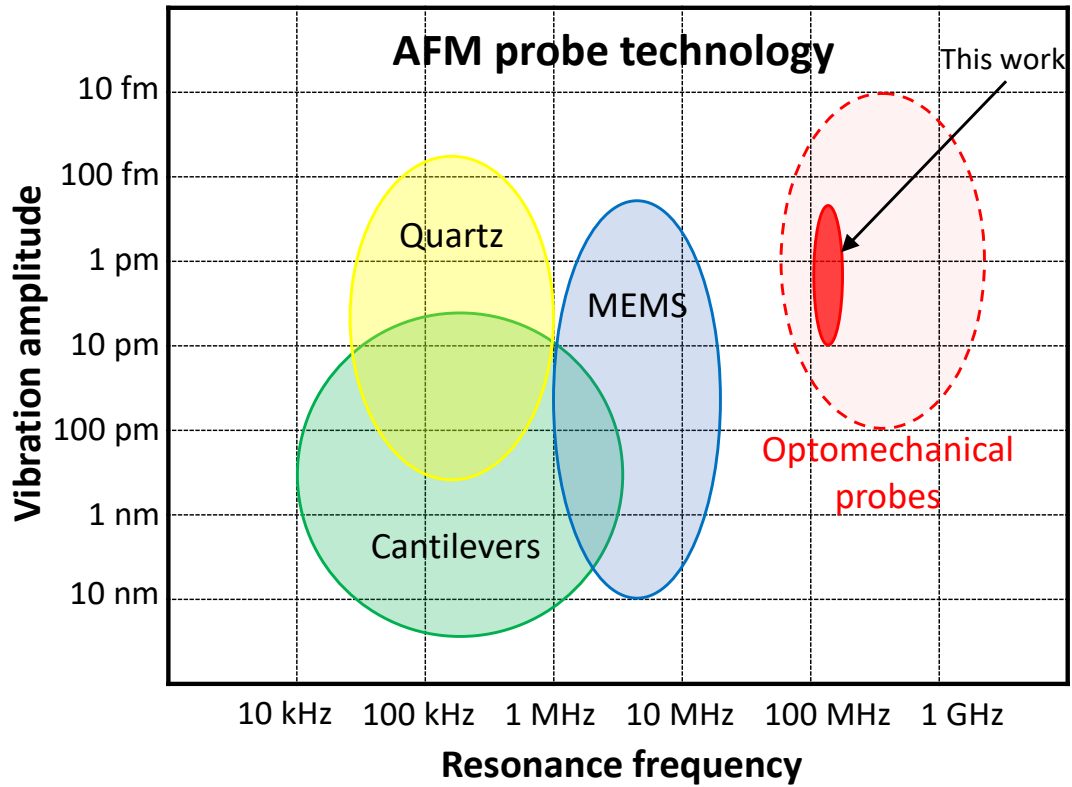

**Fig. 1: Existing AFM probe technologies for the dynamic mode.** The colored ellipses represent the typical range of vibration amplitude and mechanical resonance frequency for each probe technology. The red plain surface corresponds to the achievement of this work, while the red dashed ellipse indicates more broadly the potential of the very-high-frequency / ultra-high-frequency optomechanical technology. The lower bound of the vibration amplitude corresponds to the Brownian motion at room temperature, the upper bound to the maximum value typically met in AFM experiments.

The prevailing AFM probe, used since the beginning of the AFM age[1], has been the microfabricated cantilever, whose displacement has usually been transduced by optical beam deflection[14] or laser interferometry[15]. Its fundamental flexural mode of vibration features a low stiffness, 0.1-100 N/m, typically, which is an advantage as far as the mechanical responsivity to the measured force is concerned. The cantilever technology has benefited from numerous developments, in particular aiming at reducing its dimensions. Indeed, smaller dimensions imply higher resonance frequencies enabling high-speed applications while maintaining the desired stiffness. State-of-the-art commercially-available fast microscopes typically employ 30 µm-long cantilevers[16]. Most advanced experiments, allowing the direct observation of biomolecules in liquid at the video rate, have used even smaller cantilevers[5,9]. The vibration frequency $f$ reaches here a few MHz, while the optical detection spot faces the diffraction limit, setting a limit to the performance improvement by mere downsizing. It is interesting to note that in the last decades, many works were devoted to the integration of transducers for driving and sensing the vibration of mechanical probes, based on capacitive, piezoelectric, thermal or piezoresistive principles[17,18,19,20]. Even though these options could enable smaller cantilever dimensions and higher vibration frequencies (up to 100 MHz in ref. 20), they have not been massively adopted yet. In particular, the batch fabrication of tiny AFM cantilevers is hampered by the difficulty to produce the tips at their extremity. Another paradigm of AFM probe appeared 20 years ago, based on a quartz crystal technology. Tuning forks (TF) and length-extension resonators (LER) have been often used, especially in the field of condensed matter physics[10,21]. They offer a high stiffness, from 3 to 1000 kN/m respectively, and a highly efficient piezoelectric sensing scheme, enabling low-amplitude operation for the benefit of AFM imaging with high spatial resolution. Indeed, chemical bonding forces between two atoms have a short interaction range. They can be detected better by a low oscillation amplitude of the mechanical probe, typically below 100 pm, which improves the selectivity to the short-range signal, yielding AFM images with resolution at the scale of atoms[12,13,22]. However, TF and LER are millimetric devices whose resonance frequency $f$ ranges from a few tens of kHz to the MHz at most, which is detrimental to the measurement bandwidth that generally scales with $f$. Consequently, TF and LER are not suitable for fast AFM operation. The difficulty in bringing together short temporal resolution to investigate fast phenomena, while achieving atomic scale spatial resolution, has motivated the exploration of new concepts and architectures of AFM probes. Silicon micro-electromechanical systems (MEMS) technology offers advantages to produce AFM probes at high frequency. In the recent years, AFM operation has been achieved with probes based on in-plane vibration modes of MEMS resonators. Rings and beams have been particularly investigated which respectively embed capacitive and electrothermal/piezo-resistive transducers to actuate and sense the tip oscillation[23,24,25,26]. The in-plane geometry of the device also facilitates the fabrication of high-aspect ratio tips with nanometer apex[24,26,27]. The characteristics of these MEMS probes are represented in Figure 1. They combine a resonance frequency of up to 15 MHz and a vibration amplitude below 100 pm, extending the experimental window offered by cantilevers and quartz probes for AFM[24,26]. The limit of detection (LOD) of the vibration is at best $10^{-15}$ m/√Hz (ref. 24, 27) and the efficiency of the electromechanical transduction quickly degrades with decreasing dimensions, setting a strict limit to the maximum operation frequency reachable by such MEMS technology[28]. The LOD, being inversely proportional to the transduction efficiency, must indeed be kept below the probe's Brownian motion amplitude to benefit from optimal force resolution set at the thermomechanical limit. This constraint is all the more stringent that a higher resonance frequency generally comes along with a higher stiffness, further lowering Brownian motion amplitude.

For a bit more than a decade, the growing understanding of optomechanical interactions in semiconductor materials combined with advances in micro- and nano-fabrication have enabled the realization of miniature devices combining high quality factor ($Q_{opt}$) optical cavities and mechanical resonators on a chip[29]. In such systems, the large optomechanical coupling at play between the optical and mechanical modes allows the on-chip detection of sub-femtometer displacements[30]. It was identified that such a level of performance could impact mechanical sensing applications and in particular AFM probe technology[31,32,33]. From 2011, a flexural cantilever for AFM operation in static mode was investigated, with mechanical frequency of 4 MHz, the displacement of which was transduced by a microdisk optical cavity placed in close proximity with a LOD of $7\times10^{-15}$ m/√Hz in the linear optical regime[31,34,35]. In parallel, extreme displacement sensitivity and LOD reaching $10^{-18}$ m/√Hz were obtained even for resonators of a few micrometers in size, with resonance frequencies in the GHz range[36,37]. Force sensing in the optomechanical self-oscillation regime was also achieved[38]. Recently, we introduced the concept of an optomechanical probe for VHF sensing of force, where a silicon ring forming an optical cavity also acted as a probe mechanical resonator with a tip for force detection[39]. We demonstrated optically actuated and detected mechanical oscillations above 100 MHz of frequency, and evidenced the detection of mechanical interactions with the tip of the probe[32]. Cavity optomechanics thus appears as a promising approach to produce VHF probes for AFM operation in dynamic mode with ultra-low oscillation amplitudes, exceeding the limits of current technologies as illustrated in Figure 1. Building on our previous work, the present study takes a further step towards the actual exploitation of VHF optomechanical probes in an AFM configuration. The fabrication of the devices is achieved using very-large scale integration (VLSI) technologies inherited from those of silicon photonics. The probe is actuated using either capacitive or optical forces and a comparison between both is led. The vibration of the tip is optomechanically transduced, with a resonance frequency larger than 100 MHz. The Brownian motion is resolved with a LOD of $3.10^{-16}$ m/√Hz. A custom instrument for fast data acquisition and processing is used to get further results in an AFM configuration in air. Force curves show the probe's sensitivity to the mechanical interaction with a surface in a regime of very low vibration, where the contact is mediated by a water meniscus. In addition, stable closed-loop operation is demonstrated. These achievements lay the foundations for a new class of AFM probes enabled by cavity optomechanics, validating operation in the dynamic mode at very high oscillation frequency and ultra-low amplitude in an instrumental configuration amenable to AFM imaging.

## Results and discussion

### Principle of operation

Figure 2 gives an overview of the concept of the optomechanical AFM probe. As shown in Figure 2a, an AFM probe is basically a harmonic mechanical resonator driven by an actuation signal. Any interaction force between the probe's tip and the sample surface impacts the eigenfrequency and/or the dissipation of the resonator, which impacts in turn the amplitude and/or the phase of the vibration signal. In a reduced-order mechanical model, the probe is described by its free resonance frequency $f_0$, quality factor $Q$ and effective stiffness $K_{eff}$ (See "Methods"). Our specific optomechanical AFM probe design is illustrated in Figure 2b. It consists of a silicon ring acting both as the mechanical resonator and an optical cavity. In such a structure, the extensional mechanical modes and the optical ring modes are strongly coupled[29,40], which leads to an efficient optical transduction of mechanical motion.

The laser light traveling in the waveguide evanescently couples in and out of the ring. The detection of motion, carried through the output optical signal, is also favored by the high quality factor of the optical resonance. The device is designed to operate at a wavelength close to 1.55 μm. Actuation of the vibration of the ring can be obtained through electrostatic forces by applying a sine voltage to the electrodes in close proximity of the ring or through optical forces by modulating the intensity of the optical field injected and stored in the cavity. Figure 2c shows the shape of the extensional mechanical mode used in the present study to reach the VHF range. Its azimuthal order is 9, allowing to locate the 3 spokes at nodes and the tip at a maximum of vibration amplitude. In this configuration the probe's tip is set into a quasi-1D vertical oscillation by the vibration of the ring, which is actuated with a significant amplitude.

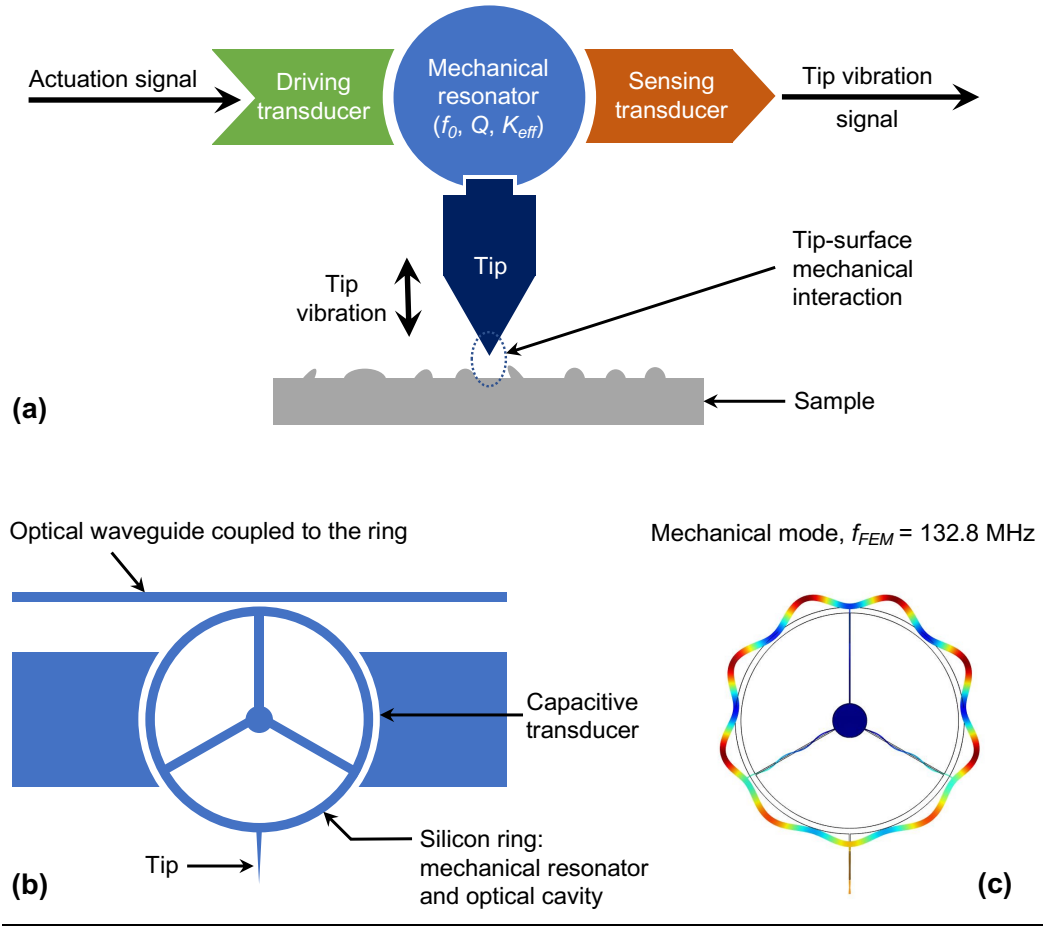

**Fig. 2: Introduction of the AFM probe with optomechanical transduction.** (a) Block diagram of the principle of operation of an AFM probe in dynamic mode. (b) Schematic of the optomechanical AFM probe (top view). The silicon ring is the mechanical resonator of the AFM probe which supports high quality factor optical modes coupled to the mechanical ones. The ring is anchored by 3 spokes. A waveguide is evanescently coupled to the optical ring cavity enabling the optomechanical transduction readout. The probe tip is diametrically opposed to the evanescent coupling region. (c) Calculation by Finite Element Modeling (FEM) of the modulus of the displacement of the mechanical mode under study in this work. A resonance frequency of $f_{FEM}$ = 132.8 MHz is obtained here for a ring of 20 μm in diameter and of 220 nm in thickness, corresponding to our design parameters.

**Implementation and fabrication**

Very large-scale integration (VLSI) clean-room techniques derived from our standard silicon photonics technologies are used to fabricate the VHF optomechanical probes on 200-mm silicon-on-insulator (SOI) wafers. The general layout and the design of the chips are presented in Figure 3a. Electrical pads located at the edges of the chip give access to the capacitive transducers of the probe. The optical waveguide coupled to the ring extends up to the center of the chip where grating couplers allow a convenient interconnection with optical fibers. A schematic overview of the fabrication procedure is depicted in Figure 3b. A first lithographic step and reactive ion etching (RIE) are used to pattern the grating couplers in the 220-nm silicon device layer of the SOI wafer. In a second step, variable shape electron beam lithography followed by a RIE step defines the main elements of the probe in the whole device layer thickness, including the ring, its anchors and its tip, the waveguide, the capacitive transducers and their associated electrical interconnects. Back-side deep RIE, or alternatively a combination of saw dicing and RIE, is realized through the whole thickness of the underlying silicon substrate in order to make the probe chips detachable from the wafer. The ring of the probe is finally released by etching the 1 µm-thick sacrificial oxide layer using vapor HF, the under-etching being controlled so that the ring becomes free to vibrate but remains anchored to the substrate through the spokes and the wider central post. Results of the fabrication process are shown in Figure 3c-e. In particular, Figure 3d shows a device with fiber transposers aligned and glued to the probe chip to ensure interconnects between the grating couplers and the optical fibers. Figure 3e gives a closer view of the optomechanical resonator of an AFM probe. The tip length is 5 µm and we estimate that the curvature radius at its apex is smaller than 30 nm (See Fig. S1 in "Supplementary Information"). The waveguide-ring optical coupling is obtained by spatial overlap of the evanescent tails of the propagative mode of the waveguide and of the mode of interest of the ring cavity. It scales exponentially with the gap distance $g$ between both objects. Total transfer is achievable at the optical resonance for the so-called critical coupling. In our devices, the critical coupling condition is typically met for a gap distance $g = 100$ nm, a distance that maximizes here the efficiency of the optomechanical transduction. This value was determined first by Finite Element Modeling (FEM) and then finely tuned by experimental characterizations. The VLSI process is well-controlled and reproducible in terms of fabrication precision, typically less than 10 nm with respect to the design, which ensures that the devices can routinely operate close to the optimal condition. The intrinsic quality factor of the optical cavity mode depends on many parameters such as the roughness of the sidewalls after fabrication or the exact design of the structure (tip and spoke positions). In our experiments we observe optical ring modes featuring intrinsic quality factors in the range of $10^4$ to $10^5$ (refer to ref. 41 for a detailed study of the optical losses).

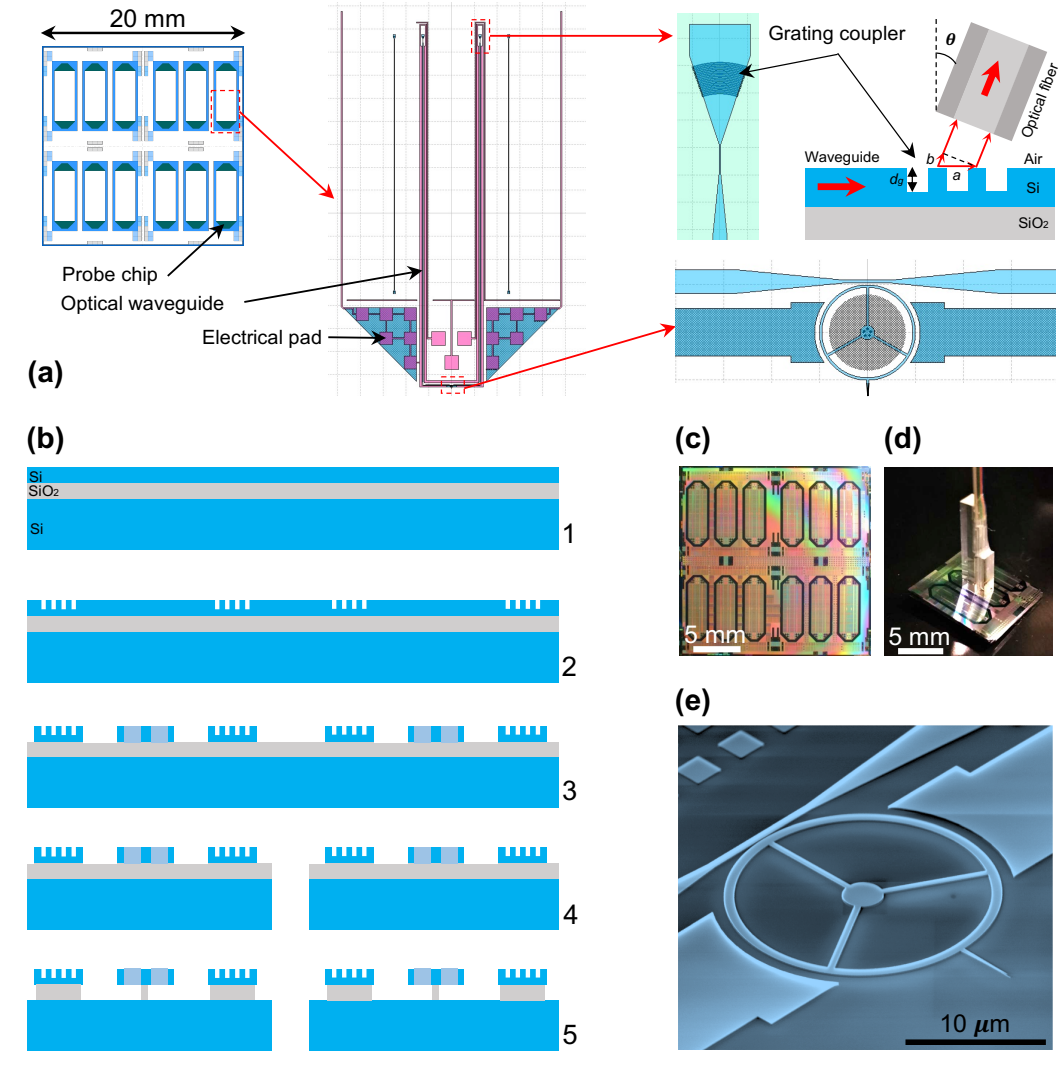

**Fig. 3: Design and fabrication of the optomechanical probe.** (a) 2 cm × 2 cm unit cells are repeated on the 200 mm wafers, each of them containing 12 probe chips. The optomechanical ring resonator holding the AFM tip is placed at the edge of the probe chip, surrounded by large pads for electrical interconnects to the capacitive electrodes. The optical waveguide extends up to the center of the probe chip. Its ends are connected to grating couplers patterned in the silicon device layer. The geometrical parameters (period *a*, height $d_g$) define the first-order diffraction angular direction $\theta$ allowing the guided light to be coupled out of plane to optical fibers. (b) Overview of the main steps of the fabrication process. 1- Photonics-grade silicon-on-insulator (SOI) wafers for the fabrication. Device layer, buried oxide layer and handle layer thicknesses are 220 nm, 1 μm and 725 μm, respectively. 2- Lithography and partial plasma etching of the device layer (etching depth = $d_g$) for the realization of the diffraction gratings. 3- Variable shape electron beam lithography and plasma etching of the device layer for the realization of all parts of the optomechanical probes. 4- Dicing of individual probe chips. 5- Releasing step by time-controlled etching of the buried oxide layer in vapor HF. (c) Optical microscopy image of a unit cell containing 12 devices at step 3 of the fabrication process. (d) View of a device with a glued transposer allowing interconnects between the grating couplers and the optical fibers. (e) Scanning electron microscopy image of a fabricated optomechanical probe. Ring diameter and width are 20 μm and 750 nm, respectively. The 3 spokes are 500 nm wide. Tip length is 5 μm. The gap distance between the waveguide and the ring is *g* = 100 nm. The distance between the capacitive transducers and the ring is 1 μm.

**Actuation and optomechanical detection**

The experimental setup for characterizations is shown in Figure 4a. The optomechanical probe can be sinusoidally driven by two mechanisms: (i) capacitively *via* electrostatic forces using the electrodes close to the ring, and optically *via* radiation pressure and photothermal effects by modulating the optical field intensity injected into the cavity[29,40]. The mechanism of optomechanical detection is schematically illustrated in Figure 4b. The mechanical vibration of the ring and tip induces a change in the optical path length of the ring cavity, such that the resonant wavelength $\lambda_{cav}$ changes in accordance, producing a change in the optical signal $T$ transmitted through the waveguide, providing that the laser is tuned to a flank of the optical resonance. As depicted in Figure 4c, we select a laser light at a wavelength $\lambda_{laser}$, blue-detuned from the peak, in the region of maximum slope, where a small change in the cavity resonance wavelength $\lambda_{cav}$ translates linearly into a maximal change of the optical transmission $\Delta T$. The light exiting the cavity and carrying information about the probe vibration is detected by a fast photodetector. The signal vector is then demodulated by a lock-in amplifier (LIA). In Figure 4d, the frequency response of the AFM probe is measured by capacitive actuation and optomechanical detection (See "Methods"). Magnitude and phase of the probe are presented for sine-wave frequency sweeps and for several values of the driving amplitude $V_{drive}$. A Lorentzian shaped resonance is observed at $f_0$ = 128.45 MHz with a quality factor $Q$ of 9000 under vacuum. The phase exhibits a quasi-ideal rotation of 180° when sweeping the frequency over the mechanical resonance. In the absence of actuation signal ($V_{drive}$ = 0 V, black line), the Brownian motion spectrum of the probe is clearly detected more than 15 dB above the noise floor of $3.10^{-16}$ m/√Hz. This indicates that the noise added to the signal is dominated by the thermomechanical noise when the resonator oscillates in the mechanical mode. When the oscillation is driven ($V_{drive}$ from 1.5 V to 10 V, color lines), the vibration amplitude of the probe tip at resonance varies from 90 fm to 13 pm depending on the value $V_{drive}$. Note that the calibration factor used to convert the output voltage of the LIA into the tip motion amplitude was determined from the Brownian noise spectrum as detailed in "Methods". In Figure 4e, we demonstrate in contrast the all-optical operation of the probe resonator. In this configuration, the incident laser light is amplitude-modulated causing the mechanical actuation of the probe resonator by the optical forces in the cavity. The Brownian motion of the probe ($V_{drive}$ = 0 V, black line) shows the same resonance frequency $f_0$ = 128.45 MHz and quality factor $Q$ = 9000 under vacuum as before. In the driven mode ($V_{drive}$ from 50 mV to 800 mV, color lines), asymmetric and shifted signals are observed in amplitude and phase, meaning that a background signal, coherent with the actuation signal, interferes with the optomechanically transduced signal of the probe vibration. This phenomenon is similar to Fano resonances[42]. It is inherent to the all-optical actuation/detection scheme using one laser: unless the critical coupling condition between the ring and the waveguide is fully met, a part of the amplitude-modulated injected light is directly transmitted to the output without entering the ring, being responsible for the background signal. As a consequence, the amplitude and phase in the mechanical domain cannot be directly read from the measurements but must instead be obtained from indirect calibration or from a model. Note that a complete model of this Fano interference has been recently developed, that embeds both photothermal and radiation pressure effects[43]. Even without full modelling here, a useful estimation of the motion amplitude can be obtained by examining the variation of the signal close to the resonance frequency: it ranges from 500 fm to 10 pm, depending on the value $V_{drive}$. Results in Figures 4d and 4e show that capacitive and optical actuations can both lead to a tip vibration amplitude in the few picometer range under vacuum. Unlike capacitive actuation, optical actuation does not require electrical interconnects to the probe chip, making the device simpler and more convenient to use for AFM purpose. The drawback is the Fano phenomenon that hinders straightforward access to the mechanical information from the optical signal. It can however be recovered by subtracting a vector signal from the probe signal aiming to cancel the background signal, which can be implemented as data post-processing or prior to demodulation at the level of the measurement chain. The latter option is actually used in the experiments of the next section.

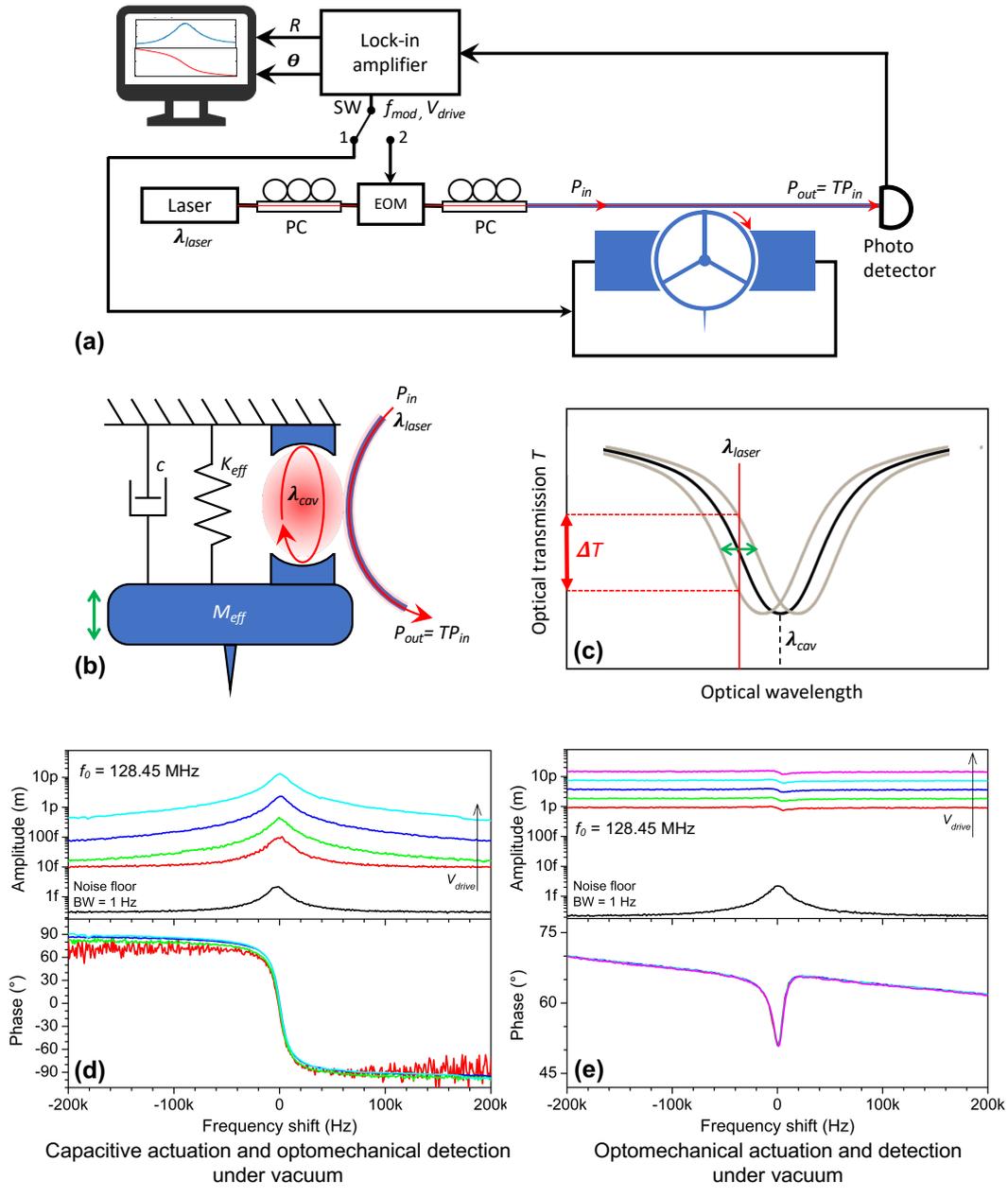

**Fig. 4: Experimental setup and results.** (a) Block diagram of the characterization set-up. A tunable laser supplies the laser light in the 1.55 μm band (wavelength $\lambda_{laser}$, output power = 2 mW). The actuation signal (amplitude $V_{drive}$, frequency $f_{mod}$) is provided by the lock-in amplifier (LIA) and is applied either to the capacitive electrodes of the device under test (DUT) (capacitive actuation, switch SW in position 1) or to the electro-optical modulator (EOM) (optical actuation, switch SW in position 2). The DUT transmitted optical signal is fed to the photodiode (PD) and demodulated in magnitude $R$ and phase $\theta$ by the LIA. PC: polarization controller. $P_{in}$: optical power at the DUT input. $P_{out}$: optical power at the DUT output. (b) Simplified diagram of the optomechanical probe coupling a mechanical resonator and an optical cavity. $M_{eff}$, $K_{eff}$, and $c$ are the mechanical parameters of the probe: effective mass, effective stiffness and damping coefficient, respectively. (c) Optical transmission spectrum of the waveguide coupled to the cavity and principle of the optomechanical transduction. The optical mode of the cavity is shown schematically in black line. The laser wavelength $\lambda_{laser}$ is blue-detuned from the optical resonance wavelength $\lambda_{cav}$. Consequently, the optical transmission varies according to $\lambda_{cav}$ (grey lines). In these conditions, the variation $\Delta T$ of the transmission is more sensitive to changes in $\lambda_{cav}$ when the quality factor of the optical mode is large. Note that the ring optical cavity supports many resonance modes (See Fig. S2 in 'Supplementary Information'). Only one is shown here for the sake of the clarity. (d) Capacitive actuation and optomechanical detection under vacuum and at room temperature. The switch SW is set in position 1. Black line: noise spectral density showing the Brownian motion of the device (resolution bandwidth = 1 Hz). The resonance frequency is $f_0$ = 128.45 MHz, $Q$ = 9000. The vibration amplitude indicated in the y-axis of the chart is calibrated from the Brownian motion considering an effective mechanical stiffness of 40 kN/m (See "Methods"). Color lines: Frequency response in magnitude and phase of the device driven using $V_{drive}$ = 1.25, 2.5, 5 and 10 V (measurement bandwidth: 1 kHz). (e) Optomechanical actuation and detection under vacuum and at room temperature. The switch SW is set in position 2. Black line: noise spectral density showing the Brownian motion of the device (resolution bandwidth = 1 Hz). The resonance frequency is $f_0$ = 128.45 MHz, $Q$ = 9000. The vibration amplitude indicated in the y-axis of the chart is calibrated from the Brownian motion considering an effective mechanical stiffness of 40 kN/m (See "Methods"). Color lines: Frequency response in magnitude and phase of the device obtained by modulating the laser power with $V_{drive}$ = 50, 100, 200, 400 and 800 mV applied to the EOM (measurement bandwidth: 1 kHz). We estimate that the optical modulation is linear with $V_{drive}$ in the range 0 to 800 mV and that the modulation is about 30% for $V_{drive}$ = 800 mV. A slope is clearly visible on the phase signal: -8 degrees over 400 kHz. It corresponds to a delay of 55 ns between the input signal and the internal reference of the LIA that we attribute to a propagation delay in the measurement chain.

**Sensing of tip mechanical interaction in the AFM configuration**

The optomechanical probes were integrated in a dedicated scanning-probe instrument, depicted in Figure 5a. It contains the same basic blocks as a standard AFM, each of them designed to match the specificities of the optomechanical probes in terms of oscillation frequency and measurement bandwidth. The probes were operated in the all-optical scheme described in the previous section for actuation and detection of mechanical motion. A signal path was added in the measurement chain, allowing to subtract a tunable vector signal from the probe signal before demodulation by a fast lock-in amplifier. This aimed at recovering the Lorentzian shape of the probe's mechanical signal by cancelling the background signal (See previous section). The demodulated signal was then fed to the processing and control unit implemented on a CompactRIO platform from National Instruments. Besides, this unit drove a Z piezo actuator that controlled the distance between the tip of the optomechanical probe and a sample. Figure 5b shows a close view of the Z piezo actuator and the optomechanical probe with its optical fibers. The Z piezo actuator was designed to offer a 100-kHz displacement bandwidth, following the work of Fleming et al.[44]. Details of the control unit are given in Figure 5c. In particular, it included: 16-bit analog-to-digital converters (ADC) for the acquisition of the probe signals at $10^6$ samples/s, a 20-bit digital-to-analog converter (DAC) for the generation of the Z piezo driving signal at $10^6$ samples/s, and a proportional-integral-derivative controller (PID) with a refresh rate of $10^6$ samples/s. The control unit allowed open-loop and closed-loop operation of the system. The delay induced by the signal processing chain of the control unit, including ADC acquisition, PID calculation, DAC and Z piezo command generation, was approximately of 3.5 µs, a value offering a phase margin large enough to allow a closed-loop bandwidth of up to 100 kHz.

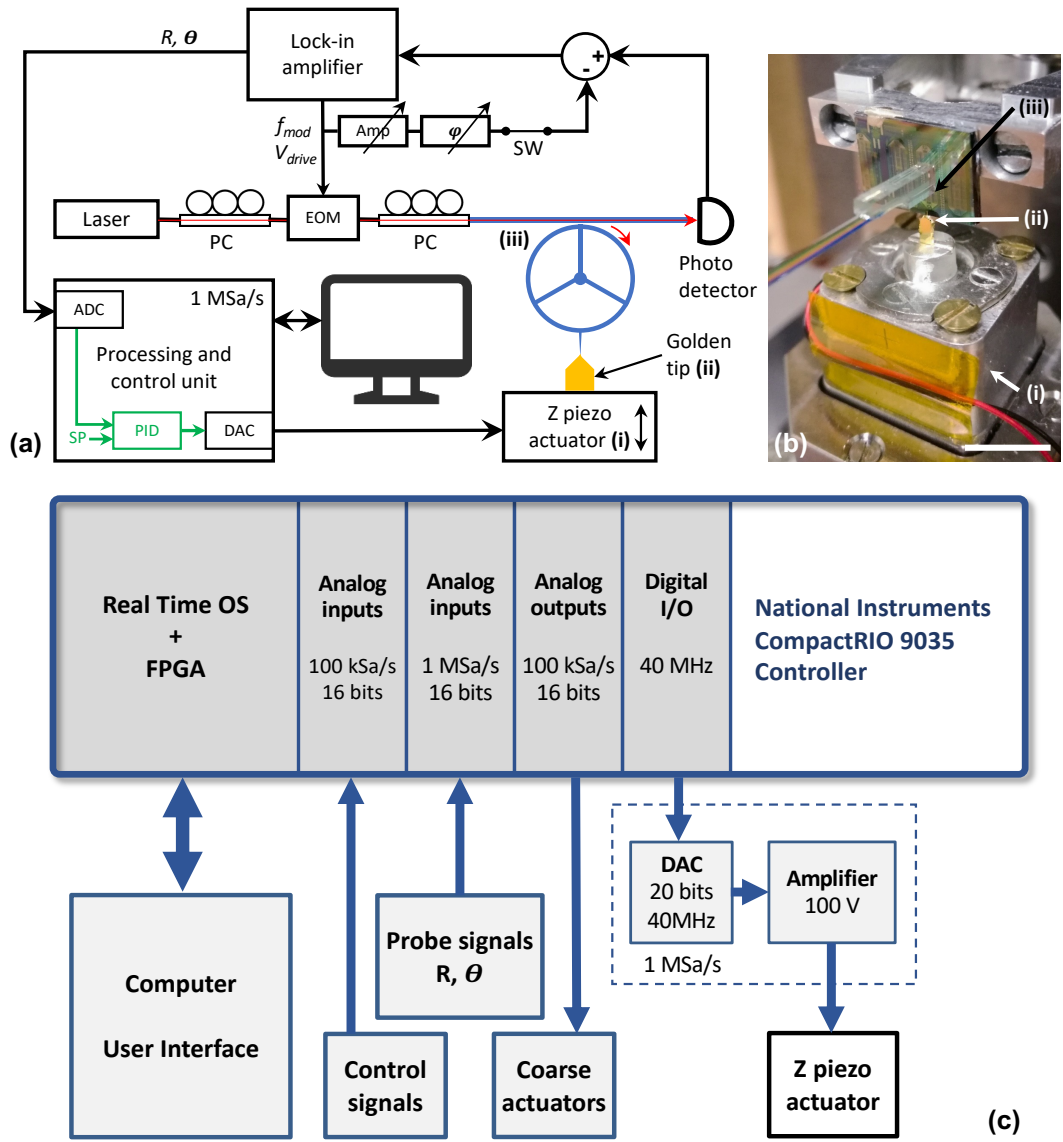

**Fig. 5: Instrument set-up for approach-retract experiments and close-loop operation.**
(a) Block diagram of the custom scanning-probe instrument used for the experiments in an AFM configuration. The optomechanical probe is optically actuated and detected. An extra signal path allows to generate a vector signal derived from the driving signal at frequency $f_{mod}$. This vector signal is tunable in amplitude (Amp) and phase (φ) and is subtracted from the probe signal before the demodulation by the fast lock-in amplifier (LIA). The demodulated information is then fed to the control unit which processes $10^6$ samples per second. The unit contains a proportional-integral-derivative controller (PID), and drives a Z piezo actuator (i) which sets the distance between the sample (ii) and the optomechanical probe (iii). PC: phase controller. EOM: electro-optical modulator. SW: switch. ADC: analog-to-digital converter. DAC: digital-to-analog converter. SP: setpoint. (b) Close view (optical image) of the AFM head, scale bar: 10 mm. The displacement range of the Z piezo actuator (i) is 2 μm, its displacement bandwidth is 100 kHz. The moving flange of the actuator holds the sample (ii), here a massive golden tip. The AFM optomechanical probe (iii) is placed vertically, its tip apex pointing towards the sample. The optical fibers are glued to the probe chip by means of a transposer aligned with the grating couplers. (c) Details of the processing and control unit implemented on a National Instruments CompactRIO 9035 controller, featured with an FPGA (field-programmable gate array), a real-time processor and embedded user interface capability. All parts of the system are programmed with the Labview software. Several input/output modules allow to acquire/generate the signals. A 20-bit DAC and a high-voltage amplifier were specifically designed for the application, offering the resolution and speed required to drive the Z piezo actuator. Extra input/output signals are available for the purpose of the control of the instrument and for a future evolution of the system.

The instrument was first used to characterize an optomechanical probe in air. The vector signal was fitted to cancel the measurement background signal. Figure 6a shows the obtained result, demonstrating that the Lorentzian shape of the mechanical information was recovered, as well as a 180-degree phase rotation at resonance. The resonance frequency of this probe was $f_0 = 130.61$ MHz and the quality factor $Q = 870$. The vibration amplitude at resonance was 0.32 pm for the drive parameters chosen in the experiment. This value is typically one order of magnitude lower than what was obtained under vacuum (Figure 4), resulting from a lower mechanical quality factor in air. In spite of a lower $Q$, the Brownian displacement noise was significantly higher than the noise floor (LOD < $3.10^{-16}$ m/√Hz) and remained the main noise contribution to the probe signal. We then performed approach-retract cycles where the sample, here a massive golden tip, was moved towards the optomechanical probe tip using the Z piezo actuator. During the experiment we recorded the piezo displacement command and the variation of the probe resonance frequency versus time, as illustrated in Figure 6b. Figure 6c represents the data as the function of the displacement command, showing more explicitly the relative frequency shift along the approach-retract cycle. Note that the frequency shift is proportional to the force gradient sensed by the optomechanical probe tip in mechanical interaction with the sample (See "Methods"). The curve displays a hysteresis cycle, typical of approach-retract experiments performed in air at a very low vibration amplitude[24]. When approaching the sample in ambient conditions, under which a water layer is generally adsorbed on the surface[12,45], the probe experiences an abrupt frequency shift due to the sudden formation of a water meniscus bridging the probe tip to the sample. It corresponds to a positive force gradient experienced by the probe, which is consistent with a spring constant added by the meniscus. The extension of the observed hysteresis is about 110 nm, corresponding to the separation distance required to break the liquid-mediated contact. Once the probe tip and surface are separated, the frequency shift recovered its initial value it had at the beginning of the approach-retract cycle where the tip was in free oscillation. The positive force gradient experienced by the optomechanical probe when the meniscus forms is estimated from the frequency shift to a value of 7 nN/nm (See "Methods"). Such value can be analyzed using the approach of Ondarçuhu et al., developed to investigate the shape and effective spring constant of liquid interfaces probed at nanometer scale[46,47]. This analysis shows that the operation of our probe with an amplitude less than the picometer causes the water meniscus to be in equilibrium at all times despite the oscillation frequency greater than 100 MHz. Moreover, we can assume that the meniscus, formed by water condensation, is laterally confined by the interaction with the probe. The meniscus spring constant can hence be approximated by[46]

$$k_{meniscus} = 2\pi\gamma/\ln(L/R) \qquad (1)$$

where $\gamma = 73$ mN/m is the surface tension of water, $R$ the radius of the probe tip and $L$ the lateral extension of the meniscus. Taking $k_{meniscus} = 7$ nN/nm and assuming a probe apex radius $R = 20$ nm, Eq. 1 implies a value $L = 180$ nm. Such a lateral extension of the meniscus yields to a reasonable description of the probe-sample interaction, consistent with the separation distance of 110 nm inferred in Figure 7 from approach-retract experiments.

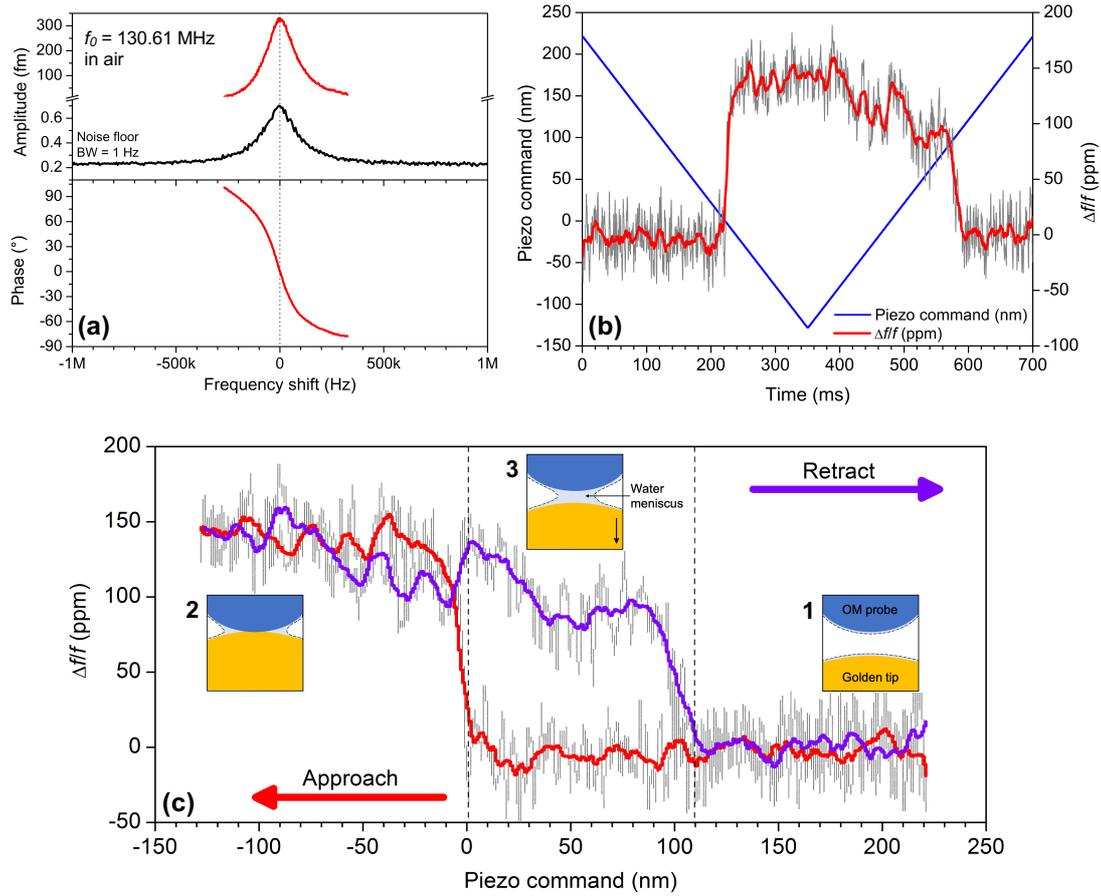

**Fig. 6: Approach-retract experiment in air.** (a) Characterization of the optomechanical probe. Black line: noise spectral density showing the Brownian motion of the device (resolution bandwidth = 1 Hz). The resonance frequency is $f_0$ = 130.61 MHz, $Q$ = 870 in air. The vibration amplitude indicated in the y-axis of the chart is calibrated from the Brownian motion considering an effective mechanical stiffness of 40 kN/m (See "Methods"). The measurement noise floor is below $3.10^{-16}$ m/√Hz. Red line: Frequency response in magnitude and phase of the device driven optically (measurement bandwidth: 1 kHz). The measurement background is cancelled by the subtraction of a vector signal at the driving frequency. Vibration amplitude is 0.32 pm at resonance. (b) Blue line: command applied to the Z piezo actuator versus time during the approach-retract experiment. Red line: relative shift of the resonance frequency of the optomechanical probe (See "Methods"). The measurement raw data are plotted in grey (acquisition rate: 1 MSa/s). Red line results from a moving average. The probe is actuated at resonance with a free vibration amplitude of 0.32 pm. (c) Variation of the relative shift of the mechanical resonance frequency versus the Z-piezo actuator command. Red line: approach of the sample towards the optomechanical probe (distance to contact is reduced). Purple line: retract from the optomechanical probe. The measurement raw data are plotted in grey (acquisition rate: 1 MSa/s). Color lines result from a moving average. The probe is driven at the resonance frequency with a free vibration amplitude of 0.32 pm. Insets 1, 2 and 3: schematic of the interaction between the optomechanical probe and the golden tip sample showing the formation of a water meniscus responsible for the hysteresis in the force curve.

**Closed-loop operation**

Closed-loop operation is a central prerequisite for the future use of the optomechanical probes for AFM imaging. It consists in controlling the probe-sample distance by means of the Z piezo actuator while maintaining the mechanical interaction constant. This was demonstrated using our custom instrument in closed-loop configuration, i.e. feeding the probe signal to the PID in order to stabilize a given setpoint value. The time response of the system is shown in Figure 7a. Before closing the regulation loop, the probe was oscillating freely at its resonance frequency with an amplitude of 0.32 pm. The resonance frequency was used as the PID input signal and the setpoint is set to a relative shift of +55 ppm, corresponding to a force gradient of +4 nN/nm. At $t = 0$, the probe tip is far from the sample and the PID commands the Z piezo actuator to reduce the distance (blue line). At $t = 90$ μs, the mechanical tip-sample interaction occurs, inducing a positive frequency shift. As described in the previous section, it corresponds to the sudden formation of a water meniscus bridging the probe tip to the sample, adding a stiffness to the probe's mechanical resonator. As the control signal exceeds the setpoint, the PID commands in turn an increase in the tip-sample distance. A stabilization stage takes place during the following 120 μs (red line) leading to a steady state (green line) where the PID input signal equals the setpoint value, and the Z piezo position remains constant. A certain time is required to reach an equilibrium where the generated meniscus yields a spring constant that sums to the probe spring to reach the targeted setpoint. Figure 7b gives a supplementary insight into the phenomenon: once the mechanical interaction is detected, the tip-sample separation is increased by 95 nm so as to reach the equilibrium state corresponding to the setpoint. If we refer to Figure 6c, this displacement is lower than the 110 nm separation needed to break the meniscus, indicating that for the equilibrium reached in closed-loop, the tip-sample interaction is mediated by the liquid meniscus. As a consequence, the Z feedback is operated with the probe tip in contact with a stretched meniscus, corresponding to the retract curve in Figure 6c (purple line). This is made possible thanks to the very low amplitude of vibration of our probe technology. In particular, such dynamic operation mode is drastically different from what occurs with cantilevers at large vibration amplitude (> 10 nm) that allow the tip to go in and out of the water meniscus when approaching the surface in ambient conditions.

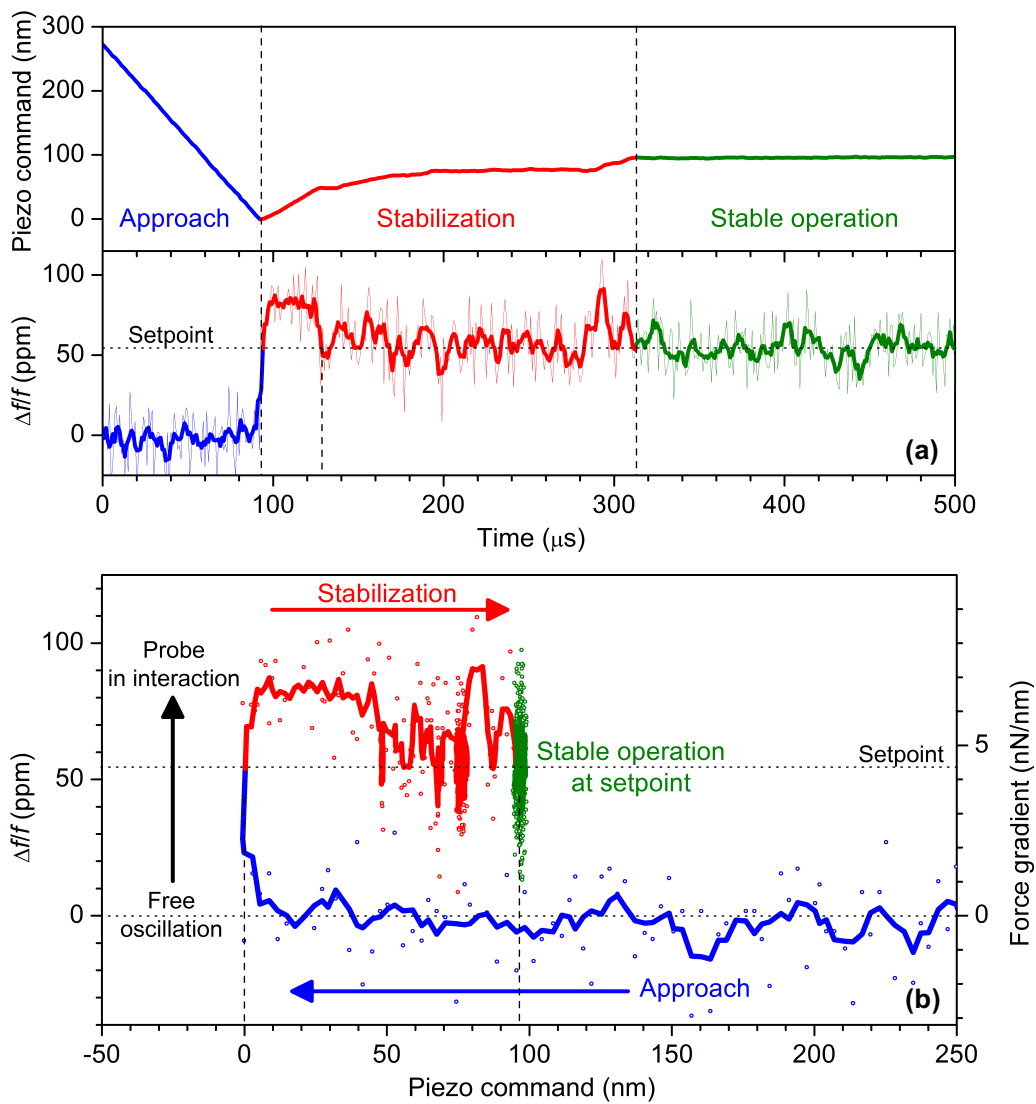

**Fig. 7: Close-loop operation of the AFM optomechanical probe.** (a) Time response of the PID signals when the feedback loop is closed. Blue line: the probe oscillates freely far from the sample, the Z piezo actuator decreases the tip-to-sample distance. Red line: the mechanical interaction occurs at $t = 90$ μs and the PID adjusts the piezo displacement to keep the signal equal to the setpoint. Green line: a steady state is reached. The Z piezo position and the probe signal (relative frequency shift) are kept stable. The measurement raw data are plotted in thin lines (acquisition rate: 1 MSa/s). Bold lines result from a moving average. (b) Variation of the probe signal (relative frequency shift) versus the piezo command when the feedback loop is closed. The chart is a representation of the signals of the time response in (a) using the same color code. The right axis represents the force gradient derived from the relative frequency shift (See "Methods").

## Conclusions

The purpose of this work was threefold: (1) to produce VHF optomechanical probes for the purpose of future AFM applications using a batch fabrication process, (2) to implement the probe sensors in an instrument in the AFM configuration, and (3) to demonstrate the capability to meet the basic requirements of AFM operation. Optomechanical probes were fabricated on 200-mm wafers using very-large scale integration (VLSI) technologies derived from the fabrication process of photonics circuits on silicon. A major feature of these probes is their very-high resonance frequency, about 130 MHz, which is greater than that of any other AFM probe technology, like short cantilevers, quartz or MEMS probes. As fabricated, the radius of curvature at the tip apex is currently of less than 30 nm. We characterized the probes under vacuum and in air, using capacitive and optical actuation, combined with optical detection. Results evidenced that the optomechanical detection enabled the observation of the Brownian motion, well above the setup noise floor lower than $3.10^{-16}$ m/√Hz. This paves the way for exquisite measurements resolutions, limited only by the thermomechanical noise at room temperature. In particular, we showed that the probes vibrate down to sub-picometer amplitudes, which is an operation regime out of reach of any other AFM technology. The probes were equipped with optical fiber interconnects and implemented in air in a dedicated instrument in the AFM configuration. Approach-retract experiments evidenced the sensitivity to the mechanical interaction with a sample, as well as the formation of a water meniscus responsible for the hysteretic behavior of the liquid-mediated contact between the probe and the sample. Moreover, we demonstrated a stable closed-loop operation of the instrument, where the probe-sample distance was adjusted to stabilize the force gradient between the tip and the sample surface. These two latter results: sensitivity to the mechanical interaction and closed-loop operation, were obtained with an ultra-low vibration amplitude of 0.32 pm. They are the central ingredients demonstrating that AFM experiments with these VHF optomechanical probes are achievable in a close future. They will be a key element for next-generation of AFM experiments, where ultra-high-speed imaging and nanosecond force tracking will be combined with a spatial resolution at the scale of the chemical bonds offered by very-low vibration amplitudes.

## Methods:

### Governing mechanics

The probe mechanical resonator is considered here as one-dimensional as shown in Figure 4d. The second-order differential equation for the lumped resonator model can be expressed as

$$M_{eff}\ddot{z} + c\dot{z} + K_{eff}z = F_{drive}(t) + F_{ts}(t) \qquad (2)$$

where $F_{drive}$ is the actuation force of the probe vibration, $F_{ts}$ is the interaction force between the tip and the sample, $M_{eff}$ is the effective mass, $K_{eff}$ is the effective spring constant, $c$ is the damping coefficient, $z$ is the tip displacement. The lumped model assumes that the total energy is conserved as well as the displacement amplitude of the probe tip, leading to the calculation of $M_{eff}$ from the shape of the vibration mode $U(r)$ obtained by Finite Element Modeling (FEM) and shown in Figure 2c:

$$M_{eff} = \frac{\rho_{Si} \int U(r)^2 dr}{U(r_t)^2} \qquad (3)$$

where $\rho_{Si} = 2330$ kg/m$^3$ is the density of silicon and $r_t$ is the position of the tip. FEM gives also the resonance frequency $f_{FEM}$, used to calculate the effective spring constant $K_{eff}$:

$$K_{eff} = \frac{M_{eff}}{(2\pi f_{FEM})^2} \qquad (4)$$

The damping coefficient $c$ can be deduced from the experimental value $Q$ of the mechanical quality factor:

$$c = \frac{2\pi M_{eff} f_{FEM}}{Q} \qquad (5)$$

**Capacitive actuation**

Capacitive actuation is obtained by applying a sine voltage $V_a(t) = V_{drive} \cos(2\pi f_{mod} t)$ to the capacitive transducers (see Figure 2b). The induced electrostatic force $F_{drive}$ acting on the mechanical resonator is proportional to the square of the applied voltage and thus oscillates at twice its frequency $f_{mod}$:

$$F_{drive}(t) = \alpha V_a(t)^2 = \frac{\alpha V_{drive}^2}{2}[1 + \cos(2\pi \times 2 f_{mod} t)] \qquad (6)$$

where $\alpha$ is a constant depending on the geometry of the capacitive transducers and on the shape of the vibration mode. Consequently, for the experiments in Figure 4d, $f_{mod}$ is swept around half the expected resonance frequency $f_0$, the probe signal being demodulated at $2 f_{mod}$ by the lock-in amplifier.

**Calibration of the vibration amplitude**

The displacements indicated in the charts in Figures 4 and 6 are calibrated from the Brownian motion spectrum, assuming the linearity of the measurement chain. The displacement noise $z_{displ\_noise}$ coming from the thermal fluctuations is given at the resonance frequency $f_0$ by[48]:

$$z_{displ\_noise} = \sqrt{\frac{2 k_B T Q}{\pi f_0 K_{eff}}} \qquad (7)$$

where $k_B$ is the Boltzmann constant and $T$ is the temperature ($T = 300$ K). It comes from Eq. 7 that $z_{displ\_noise} = 2.15$ fm at 1-Hz measurement bandwidth under vacuum (Fig. 4, $f_0 = 128.45$ MHz, $Q = 9000$, $K_{eff} = 40$ kN/m). Knowing this, displacements were plotted in Fig. 4d and Fig. 4e from the output of the LIA using calibration factors of 23.6 nm/V and 16.4 nm/V, respectively. In air, $z_{displ\_noise} = 0.65$ fm at 1-Hz measurement bandwidth (Fig. 6a, $f_0 = 130.61$ MHz, $Q = 870$, $K_{eff} = 40$ kN/m) and the calibration factor was 1.26 nm/V.

**Measurement of the relative frequency shift**

The relative frequency shift $\Delta f/f$ is obtained from the variation $\Delta\theta$ of the phase between the probe signal and the driving signal given by the lock-in amplifier. The phase variation is converted to the relative shift of the resonator eigenfrequency knowing the characteristics of the phase rotation *versus* frequency. Experiments in Figure 6b-c and Figure 7 were carried out driving the probe at its free resonance frequency, i.e. $f_{mod} = f_0 = 130.61$ MHz, where the slope *s* of the phase rotation *versus* the frequency is maximal (*s* = 836×10$^{-6}$ °/Hz in Figure 6a). Assuming small frequency shift around $f_0$, the relative frequency shift is then given by:

$$\Delta f/f = s^{-1}\Delta\theta/f_0 \qquad (8)$$

**Measurement of the force gradient**

The force gradient $\Delta k$ experienced by the probe tip is deduced from the relative frequency shift (Eq. 8) as a change of the effective spring constant of the probe:

$$\Delta k = 2K_{eff} \times \Delta f/f \qquad (9)$$

**Acknowledgments**

This work was supported by the French National Research Agency (ANR) under the research project OLYMPIA (grant ANR-14-CE26-001) and by the RENATECH French national technological network. L. Schwab acknowledges support from the *Direction Générale de l'Armement* (DGA). I. Favero acknowledges support from the European Research Council through the NOMLI project (grant 770933) and from the ANR through the Quantera QUASERT project.

## Author contributions

The probes were designed by P.E.A., I.F. and G.J., and fabricated by M.G. under the guidance of S.H. and G.J. The instrument used for the experiments in AFM configuration was conceived and fabricated by D.L, N.M., L.M., B.L. and X.D. All measurements were carried by L.S. and P.E.A., and analyzed by L.S., P.E.A., S.H., G.J., I.F. and B.L. The paper was written under the supervision of B.L.

## Conflict of interest

The authors declare that they have no conflict of interest.

Supplementary information accompanies the manuscript on the *Microsystems & Nanoengineering* website at http://www.nature/com/micronano.